\documentclass[prl,twocolumn,superscriptaddress,nobibnotes,amsmath,amssymb,showpacs]{revtex4}
\usepackage{graphics}
\usepackage[utf8]{inputenc}

\begin{document}

\title{Bimodal behavior of the heaviest fragment distribution 
in projectile fragmentation}

\author{E.~Bonnet}
\altaffiliation[present address:]{GANIL, (DSM-CEA/CNRS/IN2P3), 
F-14076 Caen cedex, France}
\affiliation{Institut de Physique Nucl\'eaire, CNRS/IN2P3,
Universit\'e Paris-Sud 11, F-91406 Orsay cedex, France}
\author{D.~Mercier}
\affiliation{LPC (CNRS/IN2P3, Ensicaen, Universit\'{e} de Caen), 
F-14050 Caen cedex, France}
\affiliation{Institut de Physique Nucl\'eaire, CNRS/IN2P3,
Universit\'e Claude Bernard Lyon 1, F-91406 Orsay cedex, France}
\author{B.~Borderie}
\affiliation{Institut de Physique Nucl\'eaire, CNRS/IN2P3,
Universit\'e Paris-Sud 11, F-91406 Orsay cedex, France}
\author{F.~Gulminelli}
\affiliation{LPC (CNRS/IN2P3, Ensicaen, Universit\'{e} de Caen), 
F-14050 Caen cedex, France}
\author{M.~F.~Rivet}
\affiliation{Institut de Physique Nucl\'eaire, CNRS/IN2P3,
Universit\'e Paris-Sud 11, F-91406 Orsay cedex, France}
\author{B.~Tamain}
\affiliation{LPC (CNRS/IN2P3, Ensicaen, Universit\'{e} de Caen), 
F-14050 Caen cedex, France}
\author{R.~Bougault}
\affiliation{LPC (CNRS/IN2P3, Ensicaen, Universit\'{e} de Caen), 
F-14050 Caen cedex, France}
\author{A.~Chbihi}
\affiliation{GANIL, (DSM-CEA/CNRS/IN2P3), 
F-14076 Caen cedex, France}
\author{R.~Dayras}
\affiliation{IRFU/SPhN, CEA Saclay, F-91191 Gif sur Yvette cedex, France}
\author{J.D.~Frankland}
\affiliation{GANIL, (DSM-CEA/CNRS/IN2P3), 
F-14076 Caen cedex, France}
\author{E.~Galichet}
\affiliation{Institut de Physique Nucl\'eaire, CNRS/IN2P3,
Universit\'e Paris-Sud 11, F-91406 Orsay cedex, France}
\affiliation{Conservatoire National des Arts et Métiers,
F-75141 Paris cedex 03, France}
\author{F.~Gagnon-Moisan}
\affiliation{Institut de Physique Nucl\'eaire, CNRS/IN2P3,
Universit\'e Paris-Sud 11, F-91406 Orsay cedex, France}
\affiliation{Laboratoire de Physique Nucléaire, Département de Physique,
de Génie Physique et d'Optique, Université Laval, Québec, Canada G1K 7P4}
\author{D.~Guinet}
\affiliation{Institut de Physique Nucl\'eaire, CNRS/IN2P3,
Universit\'e Claude Bernard Lyon 1, F-91406 Orsay cedex, France}
\author{P.~Lautesse}
\affiliation{Institut de Physique Nucl\'eaire, CNRS/IN2P3,
Universit\'e Claude Bernard Lyon 1, F-91406 Orsay cedex, France}
\author{J.~{\L}ukasik}
\affiliation{Institute of Nuclear Physics IFJ-PAN, PL-31342 Krak{\'o}w, Poland}
\author{N.~Le~Neindre}
\affiliation{LPC (CNRS/IN2P3, Ensicaen, Universit\'{e} de Caen), 
F-14050 Caen cedex, France}
\author{M.~P\^arlog}
\affiliation{LPC (CNRS/IN2P3, Ensicaen, Universit\'{e} de Caen), 
F-14050 Caen cedex, France}
\affiliation{National Institute for Physics and Nuclear Engineering,
RO-76900 Bucharest-Magurele, Romania} 
\author{E.~Rosato}
\affiliation{Dipartimento di Scienze Fisiche e Sezione INFN, Universit\`a
di Napoli ``Federico II'', I-80126 Napoli, Italy}
\author{R.~Roy}
\affiliation{Laboratoire de Physique Nucléaire, Département de Physique,
de Génie Physique et d'Optique, Université Laval, Québec, Canada G1K 7P4}
\author{M.~Vigilante}
\affiliation{Dipartimento di Scienze Fisiche e Sezione INFN, Universit\`a
di Napoli ``Federico II'', I-80126 Napoli, Italy}
\author{J.P.~Wieleczko}
\affiliation{GANIL, (DSM-CEA/CNRS/IN2P3), 
F-14076 Caen cedex, France}
\author{B.~Zwieglinski}
\affiliation{The Andrzej Soltan Institute for Nuclear Studies, PL-00681
Warsaw, Poland}
\collaboration{INDRA and ALADIN Collaborations} \noaffiliation
 
\begin{abstract}
The charge distribution of the heaviest fragment detected in the decay of 
quasi-projectiles produced in intermediate energy heavy-ion collisions has been observed to be bimodal. 
This feature is expected as a generic signal of phase transition in 
non-extensive systems. In this paper we present new analyses of 
experimental data from {Au on Au collisions at 60, 80 and 100 MeV/nucleon} 
showing that bimodality is largely independent of the data selection 
procedure, and of entrance channel effects. 
An estimate of the latent heat of the transition is extracted.
\end{abstract}

\date{\today}
\maketitle

At a first-order phase transition, the distribution of the order parameter 
in a finite system presents a characteristic bimodal behavior in the 
canonical or grandcanonical ensemble~\cite{binder,topology,zeros,noi}.
The bimodality comes from an anomalous convexity of the underlying 
microcanonical entropy~\cite{gross}. It physically corresponds to the 
simultaneous presence of two different classes of physical states for the same 
value of the control parameter, and can survive at the thermodynamic 
limit in a large class of physical systems subject to long-range 
interactions~\cite{dauxois}.
In the case of nuclear multifragmentation, a natural order 
parameter is the size of the heaviest cluster produced in each collision 
event. Indeed this observable provides an order parameter for a large 
class of transitions or critical phenomena involving complex clusters, 
from percolation to gelation, from nucleation to vaporization, 
from reversible to irreversible aggregation~\cite{botet,noi,jdf}. 

In this context, the recent observation by the INDRA-ALADIN 
collaboration~\cite{pichon}
of a sudden change in the fragmentation pattern of Au quasi-projectiles, 
loosely referred to as bimodality, has triggered a great interest 
in the heavy-ion community~\cite{wcibimo}. Looking at the correlation between the two 
heaviest fragments emitted in each event as a function of the violence 
of the collision, a clear transition is observed between a dominant 
evaporation-like decay mode, with the biggest cluster much heavier than 
the second one, and a dominant fragmentation mode, with the two heaviest 
fragments of similar size. A similar behavior has been reported in ref.~\cite{bruno}. 
Different physical scenarios have been invoked 
to interpret the phenomenon: finite-system counterpart of the nuclear 
matter liquid-gas phase transition~\cite{pichon,bonnet,gulminelli}, Jacobi 
transition of highly deformed systems~\cite{lopez_prl}, self-organized 
criticality induced by nucleon-nucleon 
collisions~\cite{trautmann,aichelin_prl}. 
In~\cite{bruno}, the two decay modes were associated to different excitation 
energies, suggesting a temperature-induced transition with non-zero 
latent heat. The qualitative agreement between refs.~\cite{pichon,bruno} 
suggests that bimodality is a generic phenomenon. 
However, differences between the two data sets subsist, and trigger 
or selection bias cannot be excluded.
To disentangle between the different scenarios, it is necessary 
to control the role of the entrance channel dynamics and
establish if the transition is of thermal character.
In this letter, to progress on these issues, event ensembles with
equiprobable excitation energy distribution are built and compared.

We present a new analysis of quasi-projectiles (QP) 
produced in Au+Au collisions measured with the INDRA apparatus~\cite{indra}
at the GSI laboratory at incident energies from 
60 to 100~MeV/nucleon~\cite{indra@gsi}. The robustness of the signal
of bimodality is tested 
against two different QP selection methods. A weighting 
procedure~\cite{gulminelli} is applied to test the independence of the 
decay from the dynamics of the entrance channel. Finally, a double 
saddle-point approximation is applied to extract from the measured data 
an equivalent-canonical distribution that can be quantitatively confronted 
to statistical theories of nuclear decay~\cite{gargi}.

In this energy regime, a part of the cross section corresponds 
to collisions with dynamical neck formation~\cite{neck}. 
We thus need to make sure that the observed change in the fragmentation
pattern~\cite{pichon} 
is not trivially due to a change in the size of the QP.
After a shape analysis in the center of mass frame~\cite{cugn}, only events with a 
total forward detected charge larger than 80\% of the Au charge were 
considered (quasi-complete events).
Two different procedures aiming at selecting events with negligible neck contribution were adopted.
In the first one~\cite{pichon} (I) by eliminating  events where the entrance 
channel dynamics induces a forward emission, in the quasi-projectile frame, 
of the heaviest fragment $Z_1$~\cite{colin}.
For isotropically decaying QPs, this procedure does not bias the event sample
but only reduces the statistics.
In a second strategy (II) the reduction of the neck contribution is
obtained by keeping only ``compact''  events by 
imposing (i) an upper limit on the relative 
velocity among fragments, and (ii) a QP size constant
within 10\%, see~\cite{bonnet} for details. 
In both cases fission events were removed~\cite{pichon}.

\begin{figure}[htbp]
\begin{center}
\resizebox{\columnwidth}{!}{\includegraphics{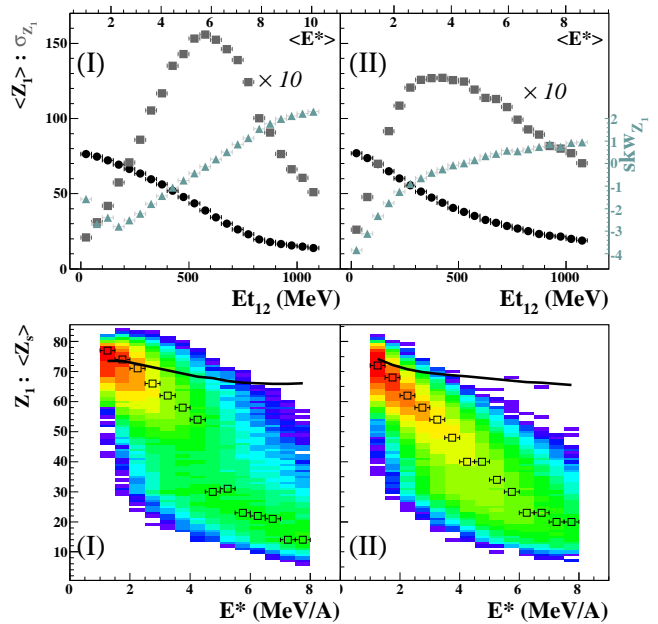}}
\caption{Upper part: average (dots), standard deviation (squares)
and skewness (triangles - right Y-axis) of the distribution of the 
heaviest fragment as a function of the light-charged particles transverse 
energy at an incident energy of 80~MeV/nucleon. Lower part: correlation
between the 
charge of the heaviest fragment and the calorimetric excitation energy. 
The open squares indicate the most probable Z1 values. 
The average total source size $Z_s$ is given by the full line.
Left side: selection (I); right side: selection (II).}
\label{fig1}
\end{center}
\end{figure}

The selected samples contain altogether about 30\% of the quasi-complete 
events at the three bombarding energies.
The main characteristics of the distribution of the heaviest fragment are 
presented in Fig.~\ref{fig1}, as a function of the total transverse energy 
of light-charged products ($Z=1,2$)~\cite{cussol}.
An excitation energy scale, estimated by calorimetry~\cite{viola}
~\footnote{To minimize the bias induced by pre-equilibrium 
emission on calorimetry, only the light-charged particles forward-emitted 
in the quasi-projectile frame are considered in the energy balance, 
and their contribution is doubled to account for backward-emitted 
particles.}, is also given.

For increasing violence of the collision, the average size of the largest 
fragment monotonically decreases. The average behavior is smooth, but 
higher moments of the distribution reveal a clear change from the 
high $Z_1$ evaporation dominated pattern, to the low $Z_1$ 
multifragmentation dominated one, passing through a region of maximal 
fluctuations where the skewness changes its sign.
These moments appear relatively independent of the selection criterion.
About one event out of four is common between the two sets; the differences 
in the observables evaluated with the two criteria thus give an estimation of the 
bias induced by the selection of data. The relative abundances observed in the correlation 
between the charge of the heaviest fragment and the deposited excitation energy are 
clearly governed by the impact parameter. 
The presence of a sudden jump in the most probable $Z_1$ value depends on the selection
method and cannot be taken as a signature of a transition, as it was proposed in previous 
works~\cite{pichon,lopez_prl,trautmann,aichelin_prl}.
The only veritable proof of bimodality would be the observation of two distinct bumps in 
the $Z_1$ distribution for a system in thermal 
contact with a heat reservoir at the transition temperature~\cite{binder,topology,zeros,noi}. 
However, the distribution of the energy deposit in a heavy-ion collision is
not determined by random exchanges with a thermal bath. This means that 
the experimental ensemble is not canonical and the $Z_1$
distribution has no meaning in terms of statistical mechanics.
To cope with this problem, a simple procedure has been proposed in 
ref.~\cite{gulminelli}. The bimodality in the canonical two-dimensional probability distribution 
$p_\beta(E^*,Z_1)$ of a system of given size $Z_s$ at a first order phase 
transition point reflects the convexity anomaly of the underlying 
density of states $W_{Z_s}(E^*,Z_1)$~\cite{binder,zeros,noi} 
according to: 
\begin{equation}
p_{\beta}(E^*,Z_1) = W_{Z_s}(E^*,Z_1) \exp (-\beta E^*) \mathcal{Z}^{-1}_{\beta},
\label{distri_cano_2D}
\end{equation}
{where $\mathcal{Z}_{\beta}$ is the partition function.} 
In an experimental sample, the energy distribution is not controlled 
by an external bath through a Boltzmann factor, but it is given by a 
collision and detector dependent functional $g(E^*)$:
\begin{equation}
p_{exp}(E^*) \propto \int dZ_1 W_{Z_s}(E^*,Z_1) g (E^*). 
\label{distri_exp}
\end{equation}
The convexity of the density of states can be directly inferred
from the measured experimental distribution, by a simple weighting of the 
probabilities associated to each deposited energy:
\begin{equation}
p_w(E^*,Z_1)=\frac{p_{exp}(E^*,Z_1)}{p_{exp}(E^*)}=\frac{p_{\beta}(E^*,Z_1)}{p_{\beta}(E^*)}=
\frac{W_{Z_s}(E^*,Z_1)}{W_{Z_s}(E^*)}.
\label{reweighting}
\end{equation}

This procedure allows to get rid of the entrance
channel impact parameter geometry that naturally favors the lower
part of the $E^*$ distribution. To produce a flat
$E^*$ distribution according to eq.(\ref{reweighting}), 
we have weighted the $Z_1$ yields in each $E^*$ bin with a factor 
proportional to the inverse of the bin statistics.

\begin{figure}[hb]
\begin{center}
\resizebox{\columnwidth}{!}{\includegraphics{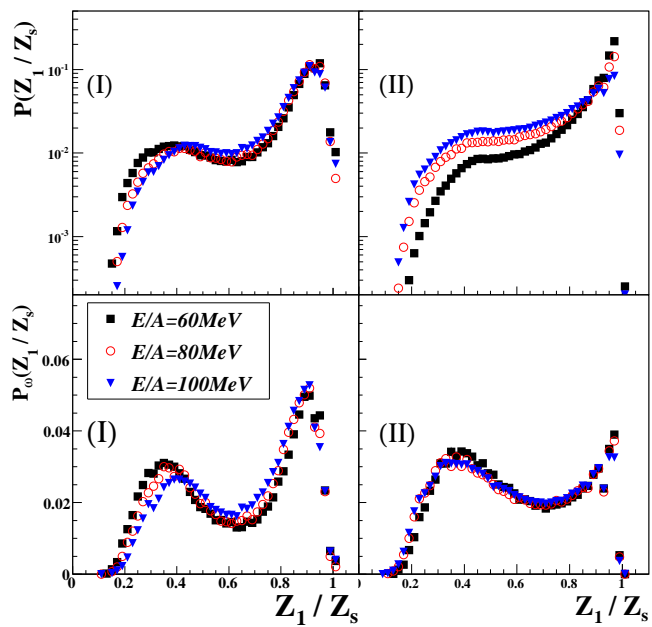}}
\caption{Upper part: measured distribution of the charge of the largest fragment
normalized to the charge of the source detected in Au+Au collisions 
at three different bombarding energies. Lower part: normalized distributions 
obtained considering the same statistics for each excitation energy bin. 
The left (right) side shows distributions obtained with the data selection 
method (I) ((II)).} 
\label{fig2}
\end{center}
\end{figure}

The results obtained with the two different selection methods are given in 
Fig.~\ref{fig2} (bottom). To take into account the small variations
of the source size, the charge of the heaviest fragment $Z_1$ has been normalized
to the source size. 
After the weighting procedure, a bimodal behavior of the largest
fragment charge clearly emerges in both cases.

Eq.(\ref{reweighting}) holds only if the bias function $g$ 
in eq.(\ref{distri_exp}) does not explicitly depend 
on $Z_1$, which is a phase-space dominance assumption.
The physical meaning of this hypothesis is that the entrance channel 
geometry and dynamics (as well as the bias induced 
by the detection system and data selection) determine only the
energy distribution and size ($Z_{s}$) of the 
QP, while for each given value of $E^*$ and $Z_{s}$ the size of the 
heaviest fragment ($Z_{1}$) is dominated by the corresponding 
available phase space.
The similarity of the two samples
at 80 MeV/nucleon, after the weighting procedure, is an
indication that the bias induced by the data sorting is small. 
The phase-space dominance hypothesis can further be checked 
by comparing the effect of the weighting procedure on data issued 
from different entrance channel dynamics. 
This is done in Fig.~\ref{fig2}, where the same weighting method 
has been applied on data at different bombarding energies. 
The comparison is not conclusive in the case of selection (I), where 
the excitation energy distributions obtained at the different incident energies 
happen to be largely superposable (Fig.~\ref{fig2} top left), and
we cannot \textit{a priori} exclude a bias function.
Conversely in the case of selection (II), we can see that the weight of the low $Z_1$ component, 
associated to more fragmented configurations and higher deposited energy, 
increases with the bombarding energy. This difference disappears when data are 
weighted, showing the validity of the phase-space dominance hypothesis.
\begin{table*}
\caption{\label{table1} Parameters of the equivalent canonical 
distribution eq.(\ref{gcano_t}) at the transition temperature as 
estimated from the two data selection methods. The $\chi^2$ of the fit 
is also given. 
}
\begin{ruledtabular}
\begin{tabular}{cccccccccccc}
 &$\rho$ & $Z_l$ & $\sigma_{Zl}$ &  $E_l$ & $\sigma_{El}$ & 
  $Z_g$ & $\sigma_{Zg}$ &  $E_g$ & $\sigma_{Eg}$ & 
  $\chi^2/N_{dof}$  \\
\hline
set (I) E/A=80MeV&-0.861 & 72.5 & 16.5 & 1.42 & 2.25 & 12.1 & 13.4  & 8.52 & 2.62 &
0.53 \\
set (I) E/A=100MeV&-0.861 & 69.3 & 15.9 & 1.67 & 2.30 & 12.1 & 13.7  & 8.76 & 2.83 &
0.59 \\
set (II) E/A=80MeV&-0.925 & 69.1 & 12.6 & 1.02 & 1.78 & 2.10 & 24.6 & 10.4 & 4.04 &
0.80 \\
set (II) E/A=100MeV&-0.925 & 68.3 & 12.5 & 1.07 & 1.77 & 2.96 & 24.4 & 10.2 & 3.96 &
0.96 \\
\end{tabular}
\end{ruledtabular}
\end{table*}

The three studied energies and the two selection criteria (I) and (II) 
produce similar but not identical distributions even after renormalization, 
meaning that a residual bias on the density of states exists.
One may ask whether this bias prevents a sorting and dynamic-independent 
extraction of the entropic properties of the system. 
To answer this question, we can compare the information on the 
coexistence zone in the $(Z_1,E^*)$ plane extracted from the different samples.
We thus have to solve eq.(\ref{reweighting}) 
for the canonical distribution $p_{\beta_t}(E^*,Z_1)$  at the transition 
temperature $\beta_t$ at which the two peaks of the energy distribution 
have the same height~\cite{noi}.
This is easily obtained in a double saddle point approximation~\cite{gulminelli}:
\begin{equation}
p_{\beta_t}(E^*,Z_1)= \sum_{i=l,g} N_i \frac{1}{\sqrt{det\Sigma_i}}
\exp\left ( -\frac{1}{2}\vec{x_i}\Sigma_i^{-1}\vec{x_i}\right).
\label{gcano_t}
\end{equation}
where $\vec{x_i}=\left (E^*-E_i,Z_1-Z_i\right )$.
$\Sigma_i$ represents the variance-covariance matrix and is related
to the entropy curvature matrix (see formulae 10, 11 and 12
of~\cite{gulminelli}).
The correlation coefficient
$\rho$=$\sigma_{{Z_1}{E^*}}$/$\sigma_{Z_1}\sigma_{E^*}$, which is one
of the parameters, was calculated
from the data at the three incident energies, before the weighting procedure and
for each selection method, on the largest validity domain i.e. 1-8
MeV/nucleon for (I) and 1-12 MeV/nucleon for (II) (see table
\ref{table1}). $\Sigma_i$ is evaluated 
at the liquid $l$ (gas $g$) solution, and
$N_i$ are the proportions of the two phases, with 
$N_l/N_g=\sqrt{det\Sigma_l}/\sqrt{det\Sigma_g}$.

The weighted experimental distribution can be fitted with the function 
$p_w(E^*,Z_1)=p_{\beta_t}(E^*,Z_1)/p_{\beta_t}(E^*)$ which, using 
eq.(\ref{gcano_t}), is an analytic function. $\rho$ being fixed, we have performed an 
8-parameter fit with the two data sets corresponding 
to the two selection procedures at the two higher bombarding energies
on the excitation energy range 2-7 MeV/nucleon; to avoid small
number effects only 2D-bins with significant statistics ($>0.5\%$ of the
corresponding $E^*$ slice) were used.
The obtained parameter values are given in table \ref{table1}.
In particular, we can estimate the latent heat of the transition of
the heavy nuclei produced as 
$\Delta E=E_g-E_l=8.1 (\pm0.4)_{stat.} (+1.2 -0.9)_{syst.}$~MeV/nucleon.
Statistical error was
derived from statistical errors on $E_l$ and $E_g$ and systematic errors
from the comparison between selections (I) and (II).
The latent heat is derived from a difference
and so the possible effect of systematic errors in 
the determination of excitation energy by calorimetry due
to detection limitations (neutrons are not detected 
nor fragment masses measured)~\cite{palluto} should be included in given
error bars. Note also that the deduced parameter values
$E_l$ and $E_g$ are outside the excitation energy range used for the fit.
\begin{figure}[hb]
\begin{center}
\resizebox{\columnwidth}{!}{\includegraphics{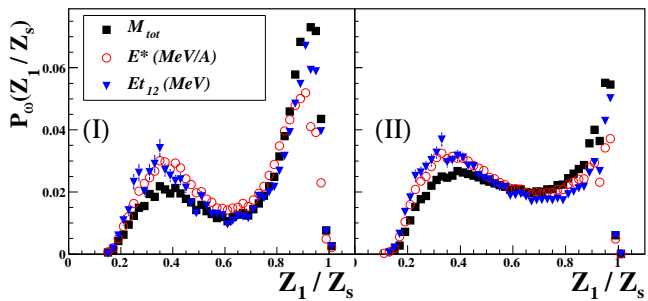}}
\caption{Experimental distribution of the largest cluster charge
normalized to have identical statistics for each excitation energy bin, 
with the two different data selection techniques (I) and (II) and for 80
MeV/nucleon incident energy. 
Different estimators of the deposited excitation energies are 
considered.} 
\label{fig3}
\end{center}
\end{figure}

Finally we use 
other estimators such as the total forward charged product 
multiplicity $M_{tot}$ and the transverse energy $Et_{12}$. 
The measured distributions weighted via eq.(\ref{reweighting}) with 
these different estimators are presented in Fig.~\ref{fig3}. We can see 
that 
bimodality is preserved in all cases, and the different energy 
estimators predict close positions for the two peaks.

To conclude,
in this paper we have presented a comparative analysis of the
quasi-projectile Au+Au data collected with the
INDRA apparatus at incident energies between 60 and 100~MeV/nucleon.
Two different methods for  quasi-projectile selection have been used,
which do not select the same physical events.  
Once the trivial
entrance channel effect of the impact parameter has been removed by
weighting the Z$_1$ distribution by the statistics of the
excitation energy distribution, a clear indication of bimodality in the 
decay pattern is observed. This behavior appears to be robust 
against the selection method, the entrance channel dynamics and the estimator of the 
deposited excitation energy.  
This analysis supports the interpretation 
of the discontinuity already observed in the decay pattern~\cite{pichon} as 
the finite system counterpart of a first order phase transition.
A multidimensional 
fit allows to extract, through a double saddle point approximation, the coexistence 
zone and a first estimate of the latent heat of the transition. 

The present results are coherent with other signals from Au
quasi-projectiles
considered indicative of a first order phase transition like a
fossil signal of spinodal fluctuations and configurational energy
fluctuations associated with negative heat capacity~\cite{thbonnet,Bor08}.
Interpretations given in~\cite{lopez_prl,trautmann,aichelin_prl} do not
register in that coherent picture. However it would be interesting to know if
those interpretations can verify the bimodality of $Z_1$ for the weighted
distribution and its independence of the incident energy 
as it is observed in that work. 

%
%

\end{document}